\documentclass[11pt]{article}
\usepackage{epsf}
\newlength{\bredde}
\def\slash#1{\settowidth{\bredde}{$#1$}\ifmmode\,\raisebox{.15ex}{/}
\hspace*{-\bredde} #1\else$\,\raisebox{.15ex}{/}\hspace*{-\bredde} #1$\fi}
\newcommand{\beq}{\begin{equation}}
\newcommand{\eeq}{\end{equation}}
\newcommand{\noi}{\vspace{12pt}\noindent}
\newcommand{\lG}{\raise.3ex\hbox{$\stackrel{\leftarrow}{G}$}}
\newcommand{\lU}{\raise.3ex\hbox{$\stackrel{\leftarrow}{U}$}}
\newcommand{\lP}{\raise.3ex\hbox{$\stackrel{\leftarrow}{{\cal P}}$}}
\newcommand{\leta}{\raise.3ex\hbox{$\stackrel{\leftarrow}{\eta}$}}
\newcommand{\lOmega}{\raise.3ex\hbox{$\stackrel{\leftarrow}{\Omega}$}}
\newcommand{\ldr}{\raise.3ex\hbox{$\stackrel{\leftarrow}{\delta^r}$}}

\def\beqn{\begin{eqnarray}}
\def\eeqn{\end{eqnarray}}

\def\gtwid{\raise.3ex\hbox{$>$\kern-.75em\lower1ex\hbox{$\sim$}}}
\def\ltwid{\raise.3ex\hbox{$<$\kern-.75em\lower1ex\hbox{$\sim$}}}

\begin{document}
\topmargin -1.4cm
\oddsidemargin -0.8cm
\evensidemargin -0.8cm
\title{\Large{{\bf Low-lying Eigenvalues of the QCD Dirac Operator\\ 
at Finite Temperature }}}

\vspace{1.5cm}

\author{~\\~\\
{\sc P.H. Damgaard$^a$, U.M. Heller$^b$, R. Niclasen$^a$} and 
{\sc K. Rummukainen$^{c,d}$}\\~\\~\\
$^a$The Niels Bohr Institute and $^c$NORDITA\\ Blegdamsvej 17\\ 
DK-2100 Copenhagen, Denmark\\~\\~\\
$^b$CSIT\\Florida State University\\Tallahassee, FL 32306-4130, USA\\~\\~\\
$^d$Helsinki Institute of Physics,\\
P.O.Box 9, 00014 University of Helsinki, Finland\\}
\maketitle
\vfill
\begin{abstract} 
We compute the low-lying spectrum of the staggered Dirac operator above and
below the finite temperature phase transition in both quenched QCD and 
in dynamical four flavor QCD. In both cases we find,
in the high temperature phase, a density with close to square root
behavior, $\rho(\lambda) \sim (\lambda-\lambda_0)^{1/2}$.  In the
quenched simulations we find, in addition, a volume independent tail
of small eigenvalues extending down to zero.  In the dynamical
simulations we also find a tail, decreasing with decreasing mass, at
the small end of the spectrum. However, the tail falls off quite
quickly and does not seem to extend to zero at these couplings. 
We find that the distribution of the smallest Dirac operator eigenvalues 
provides an efficient observable for an accurate determination of the 
location of the chiral phase transition, as first suggested by Jackson and
Verbaarschot.
\end{abstract}
\vfill
\begin{flushleft}
NBI-HE-00-12 \\
NORDITA-2000/30 HE\\
hep-lat/0003021
\end{flushleft}
\thispagestyle{empty}
\newpage

\setcounter{page}{1}

\section {Introduction}

The correlations of Dirac operator eigenvalues in QCD and related 
theories have been shown to have a fascinating relation to Random Matrix
Theory. There are two very different domains of interest here. One is
the so-called ``bulk'' of the eigenvalue spectrum of the Dirac operator,
far from both the infrared and the ultraviolet ends. The other is the
so-called ``hard edge'' at $\lambda \sim 0, ~i.e.$ the infrared end of the 
spectrum in theories with spontaneous breaking of chiral symmetry. The
relevance of Random Matrix Theory in describing eigenvalue correlations
of the Dirac operator in the bulk of the spectrum was first demonstrated
by Halasz and Verbaarschot \cite{HV}, and it has since been confirmed
by numerous lattice gauge theory studies \cite{bulk}. {}From a theoretical
point of view, these results in the bulk of the spectrum remain to be
better understood. In the other case, near $\lambda \sim 0$, the connection
between universal Random Matrix Theory results and the QCD Dirac operator 
spectrum \cite{SV,V,ADMN} is by now firmly established, and has also been
extensively checked in lattice gauge theory simulations \cite{latedge}. 
Already in the original work
of Shuryak and Verbaarschot \cite{SV} it was shown that the pertinent
chiral Random Matrix
Theory partition function coincides exactly with the effective field theory
partition function in the relevant microscopic
limit. More recently an explicit relationship between the 
universal Random Matrix Theory eigenvalue distributions
and those of the QCD Dirac operator has been established \cite{AD,OTV},
the precise link being the partially quenched chiral condensate
\cite{OTV,DS}. In this domain Random Matrix Theory is an intriguing alternative
description of exactly the same phenomena that can be derived from the
effective QCD partition function in the large-volume limit where $V \ll 
1/m_{\pi}^4$ \cite{LS}.

\noi
There exists also an interesting physical situation that forces us to
reconsider the two different domains of the Dirac operator spectrum
simultaneously. This is near the finite-temperature chiral phase transition,
where the analytical connection between the effective QCD partition function
and Random Matrix Theory breaks down even for the part of the spectrum
that is close to $\lambda=0$. This situation is readily confronted in
lattice gauge theory. Indeed, it is for staggered fermions even rigorously
proven~\cite{Tomboulis}, that chiral symmetry is restored at high
temperature. The low-lying spectrum of the Dirac operator must therefore
be quite different in the high temperature phase as compared with zero
temperature. In particular, the absence of chiral symmetry breaking
implies, through the Banks-Casher relation,
\begin{equation}
 \langle \bar\psi \psi \rangle ~=~ \pi \rho(0) ~,
\label{eq:Banks}
\end{equation}
that the density of eigenvalues at zero vanishes. This could happen either
just at that point of $\lambda=0$ (as, for example, in the free theory
where $\rho(\lambda) \sim \lambda^3$), or the spectrum could develop a gap.
The chiral phase transition occurs at the temperature $T_c$ where the
density of Dirac operator eigenvalues just reaches zero at $\lambda=0$.
If the transition is continuous, this will happen smoothly as the temperature
$T$ is increased towards $T_c$. In such a case, an important question
to settle is the precise power-law behavior of the spectral density of the 
Dirac operator right at $T=T_c$, 
because this can be related to the critical exponents
of the phase transition \cite{Z}. 
The same chiral Random Matrix Theory
that yields universal microscopic spectral correlators which exactly
coincide with those of the Dirac operator at $T=0$ can be tuned
in such a way as to just reach (multi-critical) points where \cite{ADMN2} 
\beq
\rho(\lambda) ~\sim~ \lambda^{2k}
\eeq
near $\lambda = 0$. Here $k$ is an integer labeling the multi-criticality.
At such points universality of microscopic spectral correlators still 
holds in the Random Matrix Theory
context, but there is no justification for assuming that these results
are relevant for the Dirac operator spectrum at $T=T_c$.\footnote{In 
particular, this behavior is only compatible with a continuous phase
transition. But at a more fundamental level, there is simply no longer
any relation between the chiral Random Matrix Theory ensemble and the
effective QCD partition function for temperatures $T$ that do not satisfy
$T \ll \Lambda_{QCD}$.} There is also
a schematic Random Matrix model for the finite-temperature behavior of
the Dirac operator spectrum \cite{JV}. It gives a different behavior
at $T=T_c$: $\rho(\lambda) \sim \lambda^{1/3}$. Also here there is no
physical
justification for using it in connection with the Dirac operator
spectrum at finite temperature, but it is an interesting model
that depends on just one deterministic external parameter, and we shall
therefore return to it below.

\noi
Suppose, for a moment, that the Dirac spectrum actually develops a gap
around $\lambda=0$ above $T_c$.
In Random Matrix Theory the end of a spectrum around such a
gap is referred to as a ``soft edge''. 
Generally, the (macroscopic) density
of eigenvalues near a soft edge behaves as (for $\lambda > 
\lambda_0$) \cite{Brezin}:
\begin{equation}
\rho(\lambda) ~\propto~ (\lambda - \lambda_0)^{2m+1/2}
\label{eq:se_rho}
\end{equation}
with $\lambda_0$ being the location of the edge. Here $m$ is an 
integer that labels universality classes classified by their Random Matrix
Theory potentials ($m$ parameters in the potentials must be tuned in
order to reach each class). Thus the {\em generic} behavior, without any
fine tuning, corresponds to $m=0$, which gives a square root approach to
the soft edge:
\begin{equation}
\rho(\lambda) ~\propto~ (\lambda - \lambda_0)^{1/2}
\label{eq:sqrt_rho}
\end{equation}
Random Matrix Theory actually gives a more detailed, microscopic, 
description. This arises from a blowing-up of the eigenvalue density function
around the soft
edge $\lambda_0$ with a rescaling according to the macroscopic
behavior (\ref{eq:se_rho}). For example, for the generic $m=0$ universality
class, the corresponding microscopic eigenvalue density is 
~\cite{Forrester}
\begin{equation}
\rho(\lambda) ~\propto~ X [{\rm Ai}(-X)]^2 + [{\rm Ai}^\prime(-X)]^2 ~,
\label{eq:Airy_rho}
\end{equation}
where $X \propto (\lambda - \lambda_0) N^{2/3}$ and ${\rm Ai}(x)$ is
the standard Airy function. Here $N$ denotes the size of the matrix, and
the rescaling by $N^{2/3}$ is required in order to spread out the increasing
number of eigenvalues to obtain one well-defined limiting function.
{}From the known asymptotic behavior of the Airy
functions one finds:
\begin{equation}
\rho(\lambda) \sim \cases{
 \frac{\sqrt{X}}{\pi} + {\cal O}\left(\frac{1}{X}\right) &
 \mbox{\textrm{for $X \to \infty$}} \cr
 \frac{17}{96\pi} |X|^{-1/2} \exp( -4 |X|^{3/2} /3) &
 \mbox{\textrm{for $X \to -\infty$}} \cr }
\label{eq:Airy_asympt}
\end{equation}
Thus, at the microscopic level the spectrum is not cut off sharply at
$\lambda_0$, but has an exponentially suppressed tail beyond.
Further, the square root behavior of the eigenvalue density is
modulated by wiggles corresponding to the distribution of particular
eigenvalues (smallest, second, third, etc.).
(See, for example, Fig.~\ref{fig:Airy_compare}.)

\noi
Until a few months ago, 
there existed only one low-statistics investigation of the
low-lying Dirac eigenvalue spectrum for staggered fermions in the high 
temperature phase~\cite{KLS}. It did not unequivocally 
establish the existence of a gap. For example, some low
eigenvalues were found. However, they could possibly
be attributed to ``would-be zero
modes'' from global gauge field topology, shifted away from zero by the
explicit chiral symmetry breaking of staggered fermions at finite
lattice spacing~\cite{Vink}. This is a general problem with staggered
fermions that also we must face here: 
at finite lattice spacing, the index theorem is not valid for 
staggered fermions, and
gauge field topology (whichever way one defines it on a discrete lattice)
does not give rise to exact zero modes of the staggered Dirac
operator. At $T=0$ no
trace of non-trivial gauge field topology on the lowest-lying
spectrum of staggered eigenvalues has been found \cite{DHNR}, except
in the 2-d Schwinger model at fairly small lattice
spacing~\cite{FHL}. Finite temperature,
which causes a depletion of genuine non-zero eigenvalues near 
$\lambda \sim 0$, further complicates this issue of a mix-up with 
would-be zero modes.

\noi
Very recently a study of just the
low-lying Dirac operator eigenvalues near $T_c$ has actually indicated
the presence of a gap at large lattice volumes \cite{Fetal}.
Again the statistics was rather limited, and only a small number of
the lowest-lying eigenvalues could be included (varying between 8 and 10).
The present
study will in many ways follow the same lines of thought as Ref.~\cite{Fetal},
but we shall have much larger statistics, and we shall also probe some
different aspects. In particular, we are also interested in seeing
whether quenching causes a different behavior of the smallest eigenvalues
near $T_c$ compared with dynamical fermions.
 
\noi
Last year a study of low-lying eigenvalues of the overlap Dirac operator
in quenched finite temperature gauge theories appeared~\cite{EHKN_fT}.
Overlap fermions are well suited for such an investigation since they
do not suffer from explicit chiral symmetry breaking even at finite lattice
spacing and since they have exact zero modes in topologically non-trivial
gauge fields. Ref.~\cite{EHKN_fT} found that effects
of topology persisted in the high temperature phase, although strongly
suppressed compared to the low temperature phase.
More interestingly, an accumulation of low eigenvalues 
with an apparently finite density in the infinite-volume limit was found
very near $\lambda\sim 0$ even above the (quenched) phase transition
temperature $T_c$.  The statistical properties of the 
smallest group of eigenvalues
were consistent with them being due to a dilute gas of instantons and
anti-instantons \cite{EHKN_fT}. These results have led to the speculation
that chiral symmetry might remain
broken even in the high temperature phase of quenched QCD with overlap
fermions. 

\noi
In this paper we describe high statistics investigations of the low-lying
eigenvalue spectrum of the staggered Dirac operator for SU(3) gauge group
at finite
temperature. We do this for both quenched (section~\ref{quenched}) and
dynamical QCD with four flavors of staggered fermions
(section~\ref{dynamical}). The interest in the quenched case lies primarily
in checking whether also staggered fermions, although insensitive to
topology at the gauge couplings we can investigate here, show signs of
unusual behavior of the smallest Dirac eigenvalues above $T_c$ (as was the
case for overlap fermions \cite{EHKN_fT}). Both
quenched and unquenched simulations give rise to Dirac operator spectra
that can be compared with Random Matrix Theory. In particular, if the
Dirac operator spectrum exhibits a gap, is it of the soft-edge kind of
Random Matrix Theory? Are there indications that Random Matrix Theory
provides a more accurate description of the Dirac operator spectrum
as the three-volume is increased? We shall try to answer these questions
in what follows.

\section {Quenched QCD at high temperature}
\label{quenched}

For our quenched finite temperature Monte Carlo
simulations we have used lattices
with temporal extent $N_t=4$ and up to three spatial volumes: $8^3$,
$12^3$ and $16^3$.  For $N_t=4$ the deconfinement phase transition for
SU(3) pure gauge theory with Wilson action
has been very accurately determined,
occurring at lattice gauge coupling $\beta_c=5.6925(2)$~\cite{T_c}.
It has for long been assumed that the chiral phase transition of the
quenched theory occurs at exactly this deconfinement phase transition
point of the pure gauge theory, but this has recently been challenged 
\cite{EHKN_fT}.

\noi
We now give some technical details of our simulations.
The gauge field configurations were generated with a mixture of
overrelaxation and heat bath updates, and were analyzed after every
20-th heat bath sweep. On each configuration we computed the 50
lowest-lying eigenvalues, and in some cases even more, using the variational
Ritz functional method \cite{Ritz}. Many ensembles consisted of several
thousand configurations. Gauge coupling, lattice size and statistics
of our ensembles is summarized in Table \ref{tab:q_ensemb}.

\begin{table}
\begin{center}
\begin{tabular}{c c c}
$V$ & $\beta$ & \#configurations \\
\hline
$8^3\times 4$ & 5.5 & 5931 \\
$8^3\times 4$ & 5.66 & 5544 \\
$8^3\times 4$ & 5.695 & 1769 \\
$8^3\times 4$ & 5.71 & 1145 \\
$8^3\times 4$ & 5.72 & 1426 \\
$8^3\times 4$ & 5.73 & 2021 \\
$8^3\times 4$ & 5.75 & 7029 \\
$12^3\times 4$ & 5.75 & 1933 \\
$16^3\times 4$ & 5.75 & 692 \\
$8^3\times 4$ & 5.8 & 5335 \\
$8^3\times 4$ & 5.85 & 6149 \\
$8^3\times 4$ & 5.9 & 4263 \\
$12^3\times 4$ & 5.9 & 1755 \\
$16^3\times 4$ & 5.9 &  824 \\
\hline
\end{tabular}
\end{center}
\caption[a]{Details of our quenched ensembles.}
\label{tab:q_ensemb}
\end{table}

\noi
As is well-known, the staggered Dirac operator
\beqn
\slash{D}_{x,y} &=& \frac{1}{2}\sum_{\mu}\eta_{\mu}(x)\left(U_{\mu}(x)
\delta_{x+\mu,y} - U^{\dagger}_{\mu}(y)\delta_{x,y+\mu}\right)\cr
&\equiv& \slash{D}_{e,o} + \slash{D}_{o,e} \label{staggeredD}
\eeqn
is anti-hermitian, with purely imaginary eigenvalues $i\lambda$ that come
in pairs of opposite sign. In eq.~(\ref{staggeredD})
\beq
\eta_{\mu}(x) ~=~ (-1)^{\sum_{\nu<\mu}x_{\nu}}
\eeq
are the usual phase factors for staggered fermions. 
Denoting 
\beq
\epsilon(x) ~=~ (-1)^{\sum_{\nu}x_{\nu}} ~,
\eeq 
we have also
explicitly shown how $\slash{D}$ connects {\em even} sites, i.e. those
with $\epsilon(x)=+1$, with {\em odd} sites, those with $\epsilon(x)=-1$, and
vice versa. This means that the operator $-\slash{D}^2$ is hermitian and
positive semi-definite. The sign function $\epsilon(x)$ defined above
plays the role of $\gamma_5$ in the continuum: it
anticommutes with $\slash{D}: ~\{\slash{D},\epsilon\}=0$. As
$-\slash{D}^2$ does not mix between even and odd lattice sites, we need only
compute the eigenvalues, on, say, the even sublattice. One easily sees that if
$\psi_e$ is a normalized eigenvector of $-\slash{D}^2$ with eigenvalue
$\lambda^2$, then $\psi_o \equiv \frac{1}{\lambda} \slash{D}_{o,e}\psi_e$
is a normalized eigenvector of $-\slash{D}^2$ with eigenvalue $\lambda^2$,
and non-zero only on odd sites. Moreover, as we will never encounter exact
zero modes on genuine quantum configurations, there is no difficulty with 
the above definition of $\psi_o$. We of course make use of these properties, 
and hence compute only the (positive) eigenvalues of
$-\slash{D}^2$ restricted to the even sublattice, and then take the
(positive) square root. All eigenvalues 
to be shown in the following thus have an equal number of negative
companions, of the exact same magnitude.

\noi
The spectral density of the Dirac operator is given by
\beq
\rho(\lambda) ~\equiv~ \frac{1}{V}\langle \sum_n\delta(\lambda -
 \lambda_n)\rangle ~,
\eeq
and it is readily computed numerically from our Monte Carlo simulations
by a binning of the measured eigenvalues per configuration at convenient
small intervals.

\begin{figure}
\begin{center}
\epsfxsize=0.6\textwidth
\epsffile{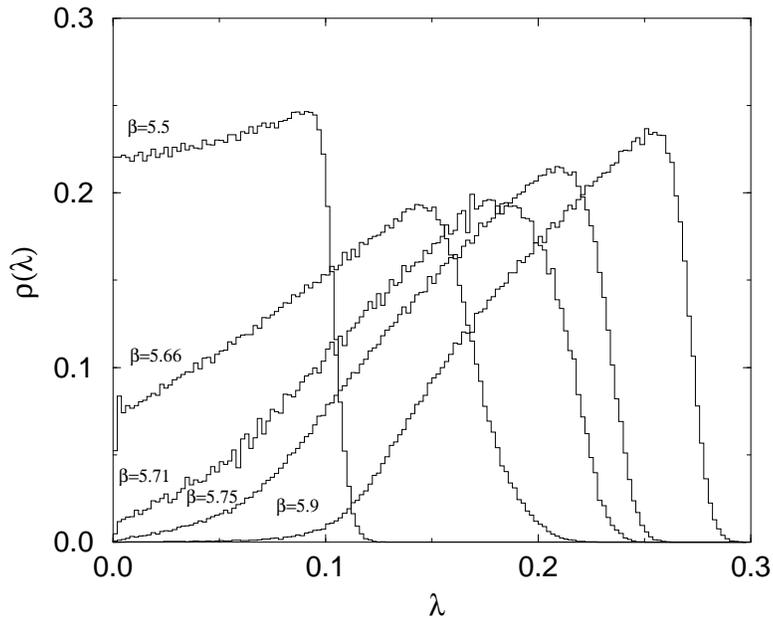}
\end{center}
\vspace*{-2mm}
\caption[a]{The measured distribution of the 50 lowest eigenvalues in quenched 
QCD at finite temperature for several $\beta$'s on $8^3 \times 4$ lattices.
One sees a clear suppression of the spectral density at the origin
$\rho(0)$ as the temperature (coupling) is increased, indicating
a diminishing chiral condensate according to the Banks-Casher formula. For
large temperatures the spectrum is heavily suppressed near the origin,
although in this quenched case a small tail always seems to extend towards
$\lambda=0$ (see below).}
\label{fig:various_quenched}
\end{figure}

\noi 
In Fig.~\ref{fig:various_quenched} we show the density of low
lying eigenvalues on $8^3 \times 4$ lattices for several $\beta$
values between 5.5 (in the confined phase) to 5.9 (in the deconfined
phase).  In each case we computed the 50 lowest positive eigenvalues
$i\lambda$ of the staggered Dirac operator. The plots are normalized
by the following condition, $\int_0^{\infty}\rho(\lambda)d\lambda=\textrm{\#eigenvalues}/V$.
In the confined phase it is quite evident that the density at zero
would be non-zero in the thermodynamic limit. As the temperature is
increased, the density at zero decreases.
There is a qualitative change of the eigenvalue density once the
temperature is increased above $T_c$. Beyond the first few eigenvalues
the eigenvalue density assumes a shape compatible with a square root
behavior (\ref{eq:sqrt_rho}). However, a sizable tail, decreasing with
increasing temperature, of small eigenvalues persists, which seems to
extend all the way down to zero.

\begin{figure}
\begin{center}
\epsfxsize=0.6\textwidth
\epsffile{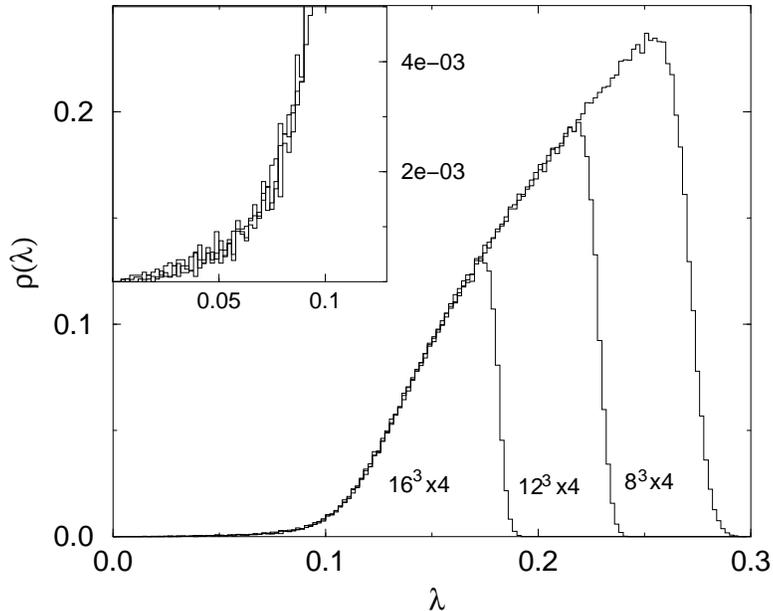}
\end{center}
\vspace*{-2mm}
\caption{The density of the smallest eigenvalues in the high
temperature phase of quenched QCD, at $\beta=5.9$,
for spatial volumes $8^3$, $12^3$
and $16^3$. In the two large volumes we measured 100 eigenvalues and
only 50 in the small volume.  Since the eigenvalue density increases
with increasing volume, the distributions are correspondingly cut off
at smaller $\lambda$. There is no observable change in the eigenvalue
density as the three-volume is increased. This holds even at the very
small tail extending towards $\lambda=0$, as shown in the magnified
plot inserted.}
\label{fig:q_vol_test}
\end{figure}

\noi
In Fig.~\ref{fig:q_vol_test} we compare the eigenvalue density in the
high temperature phase, at $\beta=5.9$, for several spatial volumes.
As can be seen the eigenvalue density is volume independent to a
surprising degree. In particular, the tail of small eigenvalues is
seen to be volume independent and appears to persist in the thermodynamic
limit. We note that the tail is much larger than would be compatible with
the exponential tail from the Airy function behavior of random matrix
theory at a soft edge eq.~(\ref{eq:Airy_asympt}) (see
Fig.~\ref{fig:Airy_compare} in section 3 below for an example).

\noi
It is tempting to speculate that the tail seen here in the quenched case
with staggered fermions is a reflection of the accumulation of small
eigenvalues seen with overlap fermions in \cite{EHKN_fT} and attributed
there to a dilute gas of instantons and anti-instantons. In the staggered
case, the explicit chiral symmetry breaking at finite lattice spacing
shifts the small modes, presumably resulting in the observed tail.

\begin{figure}
\begin{center}
\epsfxsize=0.6\textwidth
\epsffile{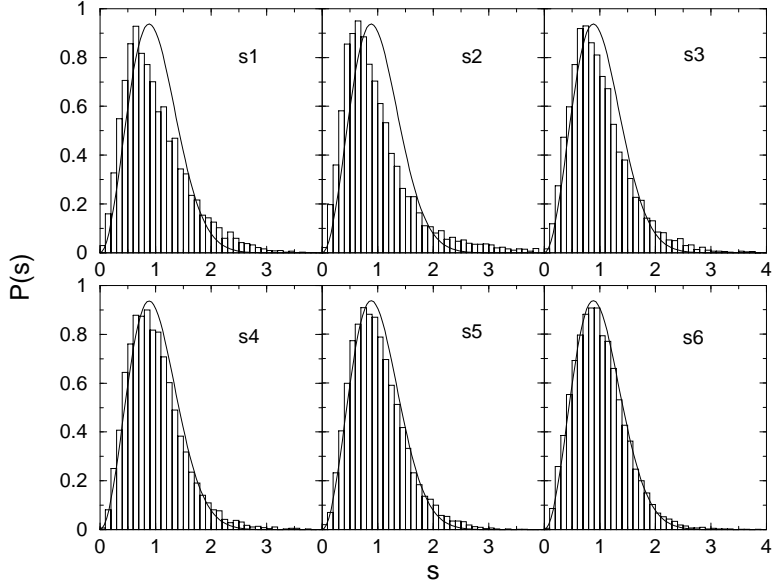}
\end{center}
\vspace*{-2mm}
\caption{Distributions of the unfolded level spacings between the
first few eigenvalues for $\beta=5.75$ on an $8^3\times 4$ lattice,
compared to the Wigner surmise.
The agreement with Random Matrix Theory is clear once we go beyond
the first few eigenvalues. However, there is not very good agreement
for the first 2-3 eigenvalues at this volume. }
\label{fig:q_level_space}
\end{figure}

\noi
We also looked at the unfolded level spacing between 
individual eigenvalues $i$ and $i+1$
\beq
s_i~=~\frac{\lambda_{i+1}-\lambda_i}{\langle\lambda_{i+1}-\lambda_i\rangle}\,.
\label{eq:level_space}
\eeq
In Fig.~\ref{fig:q_level_space} the distribution of the $s_i$ between 
eigenvalue $i$ and $i+1$, $i=1..6$ is shown and compared to the expected 
distribution, the Wigner surmise
\beq
P(s) = \frac{32}{\pi^2}s^2e^{-\frac{4}{\pi}s^2}\,.
\eeq
Evidently, the level spacings between the first eigenvalues are not very 
accurately given by Random Matrix Theory correlations. 
But as we move into the bulk, say for $i>5$ in the case of the $8^3\times 4$
lattice, the agreement with Random Matrix Theory becomes almost perfect.
It should be stressed here that this is a volume-dependent statement.
For larger volumes one needs to go beyond more eigenvalues starting at
the soft edge before one finds good agreement. This is consistent with the
fact that there is a definite scale around the soft edge, where eigenvalue
correlations are poorly described by Random Matrix Theory. Going to larger
volumes simply forces more eigenvalues into that region.

\section {QCD with four dynamical fermions at high temperature}
\label{dynamical}

For our dynamical simulations we chose to work with $n_f=4$ flavors.
Four is the ``natural'' number of flavors for staggered fermions
in the continuum limit, and an efficient exact simulation algorithm
can be used, the Hybrid Monte Carlo algorithm. In addition, the
four flavor theory is known to have a rather strong first order finite
temperature phase transition~\cite{Brown}. Therefore, tunnelings
into the low temperature phase are strongly suppressed already on
rather small systems and at temperatures quite close to $T_c$.

\begin{figure}
\begin{center}
\epsfxsize=0.6\textwidth
\epsffile{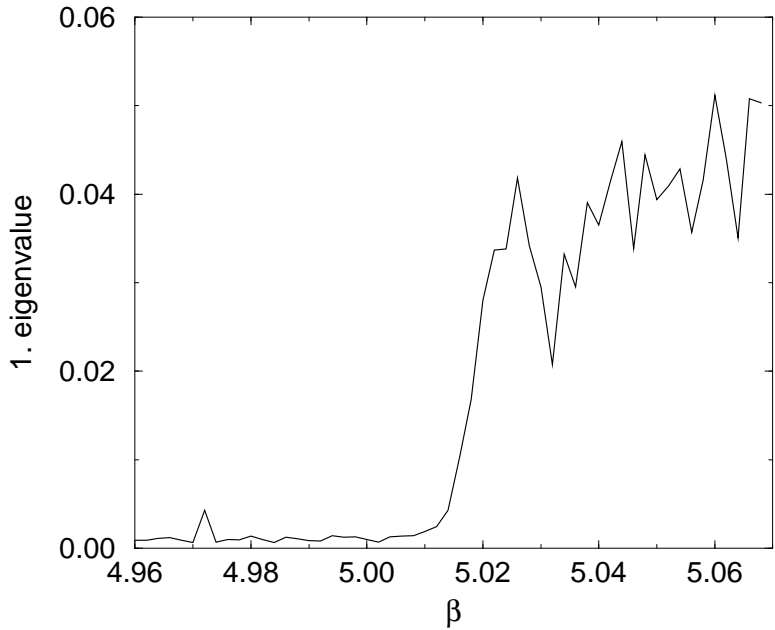}
\end{center}
\vspace*{-2mm}
\caption[a]{The lowest eigenvalue, averaged over 5 configurations, on an
$8^3 \times 4$ lattice with $am_q=0.025$ as a function of $\beta$.
One sees a clear change in behavior around $\beta \sim 5.02$, indicating
the restoration of chiral symmetry around this temperature. Just this first
Dirac operator eigenvalue can thus serve as an excellent indicator of
the chiral phase transition point. }
\label{fig:1.eig_vs_beta}
\end{figure}

\noi
We made dynamical simulations with quark masses $am_q = 0.1$, 0.05,
0.025, 0.01 and 0.002, and for couplings
$\beta$ in the high temperature phase (here $a$ is the lattice
spacing). 
Interestingly, the lowest eigenvalue provides a sensitive method for
determining the critical coupling $\beta_c(am_q)$:
in Fig~\ref{fig:1.eig_vs_beta} the lowest eigenvalue on a 
$8^3\times4$ lattice with $am_q=0.025$ is plotted against $\beta$.
$\beta$ ranges from 4.96 to 5.07 in intervals of 0.002.
Each point is an average over 5 configurations at that $\beta$-value.
A rise is observed between $\beta=5.012$ and $\beta=5.022$, which can
be clearly interpreted as the chiral symmetry restoring finite
temperature phase transition.
(For published values of the critical couplings $\beta_c(am_q)$)
see Ref.~\cite{Brown}.) The possibility of using the magnitude of the
smallest Dirac operator eigenvalues as probes for chiral symmetry
restoration was suggested by Jackson and Verbaarschot \cite{JV} on the
basis of a similarly observed behavior in a Random Matrix Theory context
that we will also discuss below. 

\noi
Most of our analysis with dynamical fermions was performed
at $\beta=5.2$, which is above $\beta_c$ for all values of $m_q$
we used.  Typically, we analyzed $\sim 3000$ configurations for each
volume and $m_q$, extracting the 50 smallest eigenvalues. Ensemble details
are summarized in Table \ref{tab:d_ensemb}. Note the large statistics
gathered for $\beta=5.2$ and $am_q=0.025$.

\begin{table}
\begin{center}
\begin{tabular}{c c c c}
$V$ & $\beta$ & $am_q$ & \#configurations \\
\hline
$8^3\times 4$ & 5.1 & 0.1 & 8897 \\
$8^3\times 4$ & 5.1 & 0.025 & 2909 \\
$8^3\times 4$ & 5.1 & 0.01 & 4050 \\
$8^3\times 4$ & 5.2 & 0.1 & 4348 \\
$8^3\times 4$ & 5.2 & 0.05 & 2149 \\
$8^3\times 4$ & 5.2 & 0.025 & 29632 \\
$12^3\times 4$ & 5.2 & 0.025 & 13929 \\
$8^3\times 4$ & 5.2 & 0.01 & 4050 \\
$12^3\times 4$ & 5.2 & 0.01 & 2882 \\
$8^3\times 4$ & 5.2 & 0.008 & 4200 \\
$8^3\times 4$ & 5.2 & 0.005 & 7948 \\
$8^3\times 4$ & 5.2 & 0.002 & 6950 \\
$8^3\times 4$ & 5.3 & 0.025 & 4549 \\
$8^3\times 4$ & 5.4 & 0.025 & 3746 \\
\hline
\end{tabular}
\end{center}
\caption[a]{Details of our ensembles with dynamical fermions.}
\label{tab:d_ensemb}
\end{table}

\begin{figure}
\begin{center}
\epsfxsize=0.6\textwidth
\epsffile{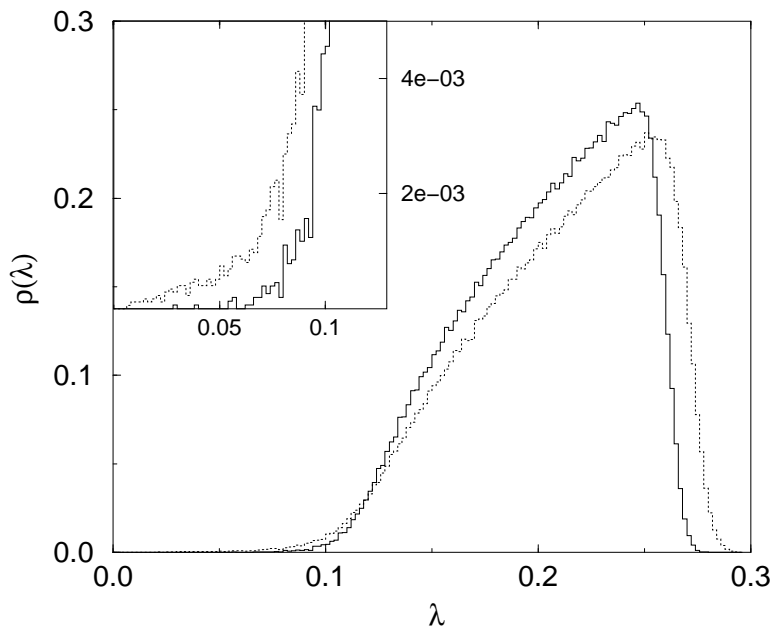}
\end{center}
\vspace*{-2mm}
\caption{Quenched spectrum of $\beta=5.9$ and a dynamical spectrum at
$\beta=5.4$ with $am_q=0.025$, both on an $8^3 \times 4$ lattice.
A blowup of the y-axis near the end of the tail
is also shown. There is a much stronger suppression of small eigenvalues
in the case of dynamical fermions, perhaps due to fewer would-be zero
modes in topological non-trivial gauge field configurations.}
\label{fig:dyn_que_compare}
\end{figure}

\begin{figure}
\begin{center}
\epsfxsize=0.6\textwidth
\epsffile{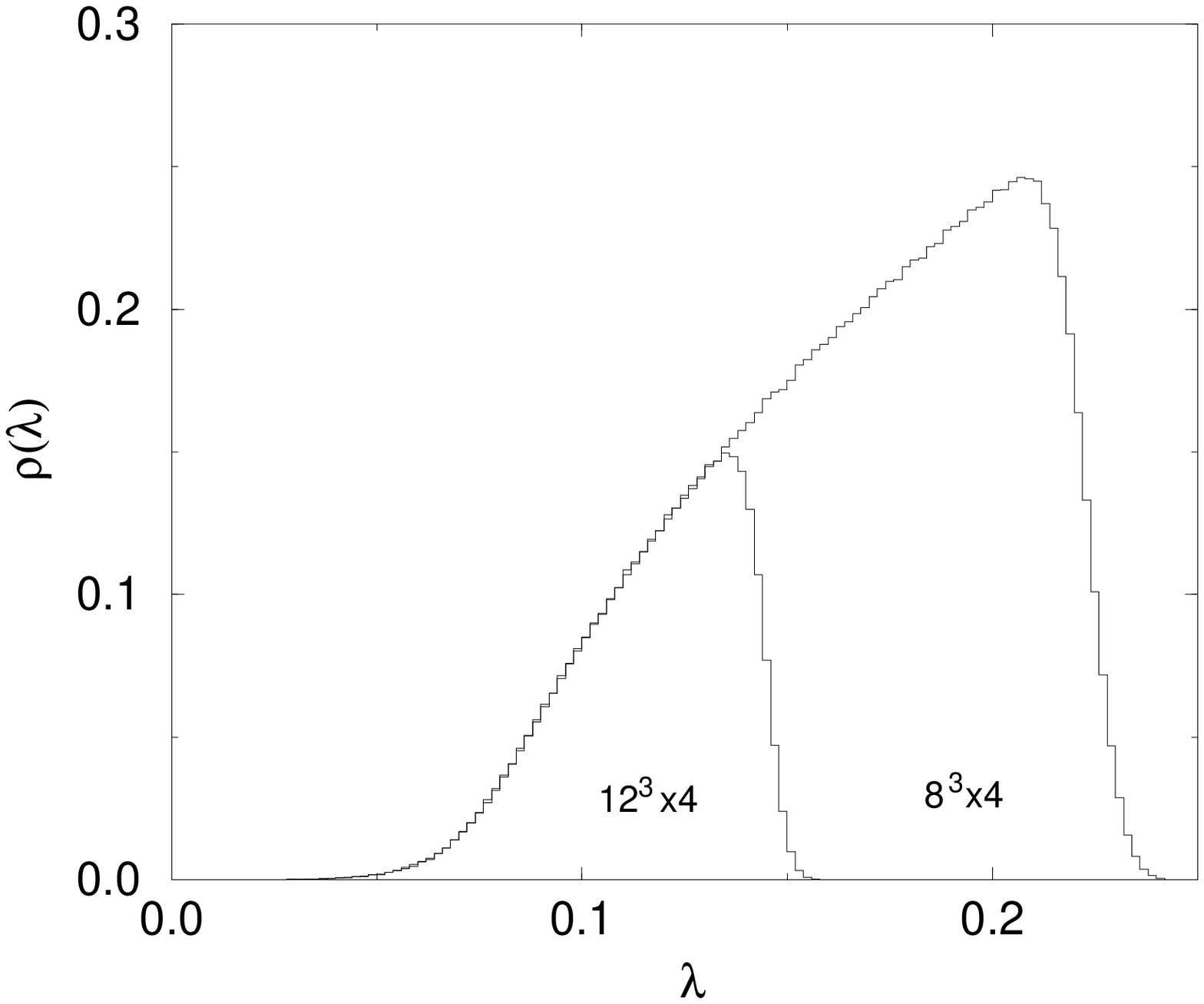}
\end{center}
\vspace*{-2mm}
\caption{The density of the 50 smallest eigenvalues in the high temperature
phase, $\beta=5.2$ and $am_q=0.025$ for spatial volumes $8^3$ and $12^3$.
The normalization of the distributions is as in Fig.~\ref{fig:q_vol_test}.
As in the quenched simulations, we observe no significant change in the 
eigenvalue spectrum at all as the three-volume is increased.}
\label{fig:d_vol_test}
\end{figure}

\noi
As in the quenched case we find a small tail at the
lower end of the eigenvalue distribution, reaching towards $\lambda=0$
from the main bulk of the distribution.   This tail also appears to be
volume independent (see Fig.~\ref{fig:d_vol_test}).  However,
in this case it is somewhat more suppressed than in the quenched case,
and for small values of the quark mass it does not extend all the way 
to $\lambda=0$, see Fig.~\ref{fig:dyn_que_compare}.

\begin{figure}
\begin{center}
\epsfxsize=0.6\textwidth
\epsffile{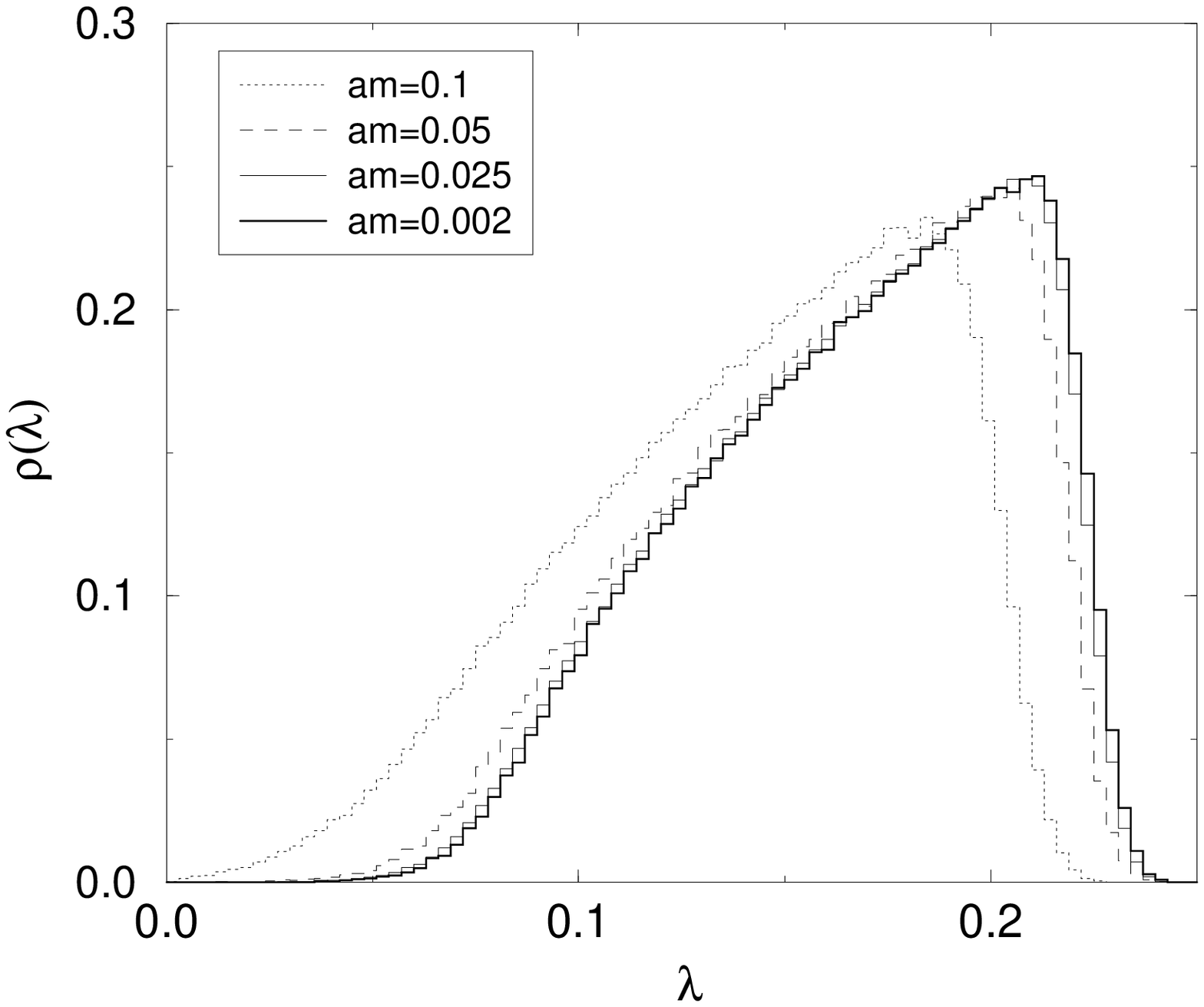}
\end{center}
\vspace*{-2mm}
\caption{The density of the 50 smallest eigenvalues at $\beta=5.2$ for quark 
masses $am_q=0.1, 0.05, 0.025, 0.002$ on $8^3 \times 4$ lattices.
While there is some shift with decreasing quark mass for the larger values
of $am_q$, the spectral density seems to approach a limiting function, which
we can identify as the chiral limit of the spectrum. Note that the 
large-mass behavior of the spectrum is almost identical to the quenched
spectrum in this region, in agreement with the notion that on this scale
of eigenvalues the large-mass fermion has completely decoupled,
and effectively become quenched.}
\label{fig:various_dynamical}
\end{figure}

In Fig.~\ref{fig:various_dynamical} we show
eigenvalue distributions measured at $\beta=5.2$ and various $am_q$.
Note that for larger $am_q$, $\beta_c$ is also larger,
which causes a shift of the whole spectrum as $am_q$ is varied.
Therefore, a direct quantitative
comparison of the distributions in Fig.~\ref{fig:various_dynamical}
is not straightforward. 
However, we note that the distributions at  $am_q=0.025$ and $am_q=0.002$
are almost on top of each other, indicating that these distributions
are very close to the $am_q=0$ distribution.

\begin{figure}
\begin{center}
\epsfxsize=0.6\textwidth
\epsffile{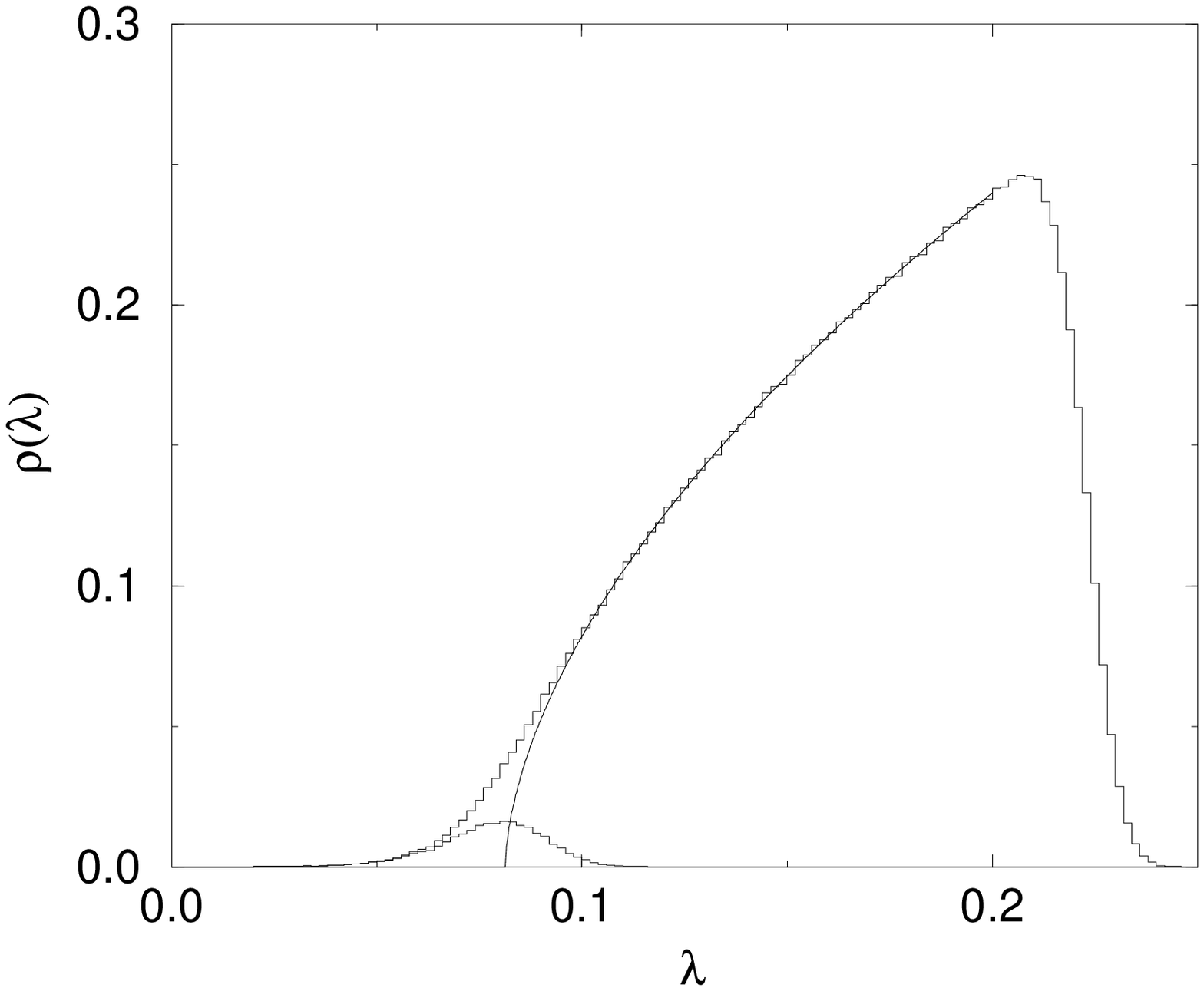}
\end{center}
\vspace*{-2mm}
\caption[a]{Power fit to the spectral density for $\beta=5.2$ with $am_q=0.025$
on an $8^3 \times 4$ lattice with cut at 0.1 (see Table~\ref{tab:fit_vs_cut}). 
Also included is the distribution of the first eigenvalue. The distribution
of this smallest eigenvalue is very different from that demanded by
the Airy-function behavior (see below).}
\label{fig:sqrt_fit}
\end{figure}

\noi
Furthermore, as in the quenched case described in the previous section, 
we find the ``macroscopic'' behavior
of the eigenvalue density seemingly compatible with a square root form.
By fitting eq.~(\ref{eq:se_rho}) to the bulk of the spectrum
(leaving out the tail), we obtain different powers, depending 
on how much of the tail we choose to cut off. This fitting has been 
done for $V=8^3\times 4$ at $\beta=5.2$ and $am_q=0.025$ and the 
resulting values can be seen in Table~\ref{tab:fit_vs_cut}.
\begin{table}
\begin{center}
\begin{tabular}{c c c}
Cut at & $m$ & $2m+1/2$\\
\hline
0.09  & 0.055 $\pm 0.003$ & 0.609 $\pm 0.006$ \\
0.095 & 0.049 $\pm 0.004$ & 0.598 $\pm 0.007$ \\
0.1   & 0.040 $\pm 0.005$ & 0.580 $\pm 0.010$ \\
0.105 & 0.036 $\pm 0.006$ & 0.572 $\pm 0.011$ \\
0.11  & 0.041 $\pm 0.008$ & 0.582 $\pm 0.016$ \\
0.115 & 0.035 $\pm 0.010$ & 0.570 $\pm 0.019$ \\
0.12  & 0.043 $\pm 0.013$ & 0.585 $\pm 0.027$ \\
0.125 & 0.040 $\pm 0.016$ & 0.581 $\pm 0.032$ \\
0.13  & 0.033 $\pm 0.021$ & 0.567 $\pm 0.041$ \\
0.14  & 0.027 $\pm 0.034$ & 0.554 $\pm 0.068$ \\
\hline
\end{tabular}
\end{center}
\caption[a]{The dependence of the fitted power on the
$\lambda$ where the small-$\lambda$ tail is cut off.
The data are from $8^3\times 4$, $\beta=5.2$ and $am_q=0.025$
lattices, using the 50 smallest eigenvalues. The fits are almost compatible
with a square-root behavior of the spectrum at this soft edge, but there
is a consistent small upward shift in the exponent.}
\label{tab:fit_vs_cut}
\end{table}
There is a clear tendency towards a slightly higher power ($=2m+1/2$)
than the square root which would be expected from Random Matrix Theory.
In Fig.~\ref{fig:sqrt_fit} we show the fit with the cut at 0.1.
For comparison, we include the distribution of the first eigenvalue in 
the figure.  Clearly, the tail at small $\lambda$ is caused by the
excessive width of the distribution of the smallest eigenvalue.

\begin{figure}
\begin{center}
\epsfxsize=0.6\textwidth
\epsffile{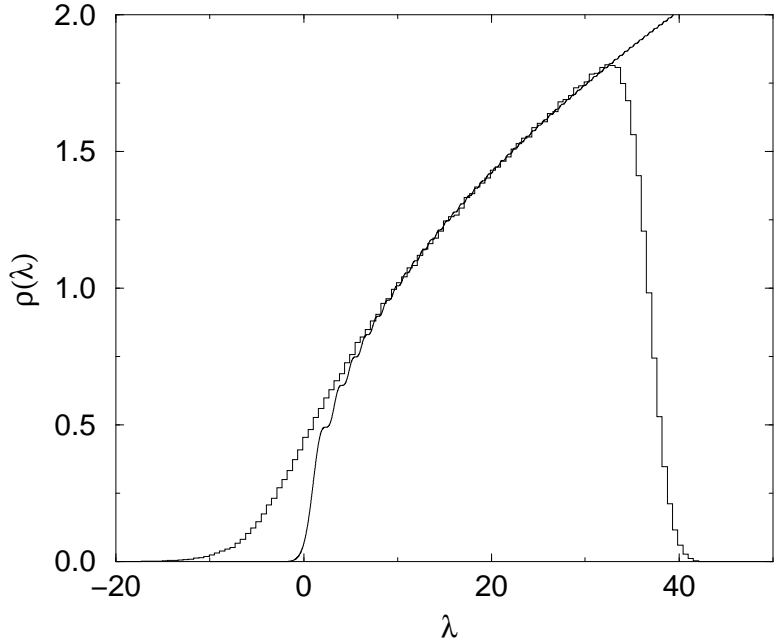}
\end{center}
\vspace*{-2mm}
\caption[a]{Comparison of the spectral density for $\beta=5.2$, $am_q=0.025$ 
on a $8^3\times4$ lattice, with the Random Matrix Theory prediction
for a soft edge, as in eq.~(\protect\ref{eq:Airy_rho}).}
\label{fig:Airy_compare}
\end{figure}

\noi
In Fig.~\ref{fig:Airy_compare} we compare the spectral density for
quark mass $am_q=0.025$ with the prediction
(\ref{eq:Airy_rho}) of the Random Matrix Theory for the density near
a soft edge.  In order to make the comparison possible, we rescale 
the spectral density as
\beq
\rho(\lambda)\rightarrow \frac{\lambda_0}{2}\left(\frac{2}{\pi\lambda_0 KV}
\right)^{\frac{2}{3}}\rho\left(\frac{2}{\lambda_0}(\lambda-\lambda_0)\left(
\frac{2}{\pi\lambda_0 KV}\right)^{-\frac{2}{3}}\right)
\eeq
where we determine $\lambda_0$ and $K$ from a fit to a square root:
$\sqrt{\frac{\lambda_0}{2}}K\sqrt{(\lambda-\lambda_0)}$.  Here $V$ is
the lattice volume.  We see that the tail predicted by the Random
Matrix Theory is much more strongly suppressed than the measured
distribution.  Moreover, we can also observe that the density of the
eigenvalues themselves does not match: the measured distribution
includes 50 eigenvalues, whereas the function (\ref{eq:Airy_rho}) has
only 40 `wiggles' in the $\lambda$-range of the distribution.
However, if we attempt to perform the comparison by matching the
number of eigenvalues/wiggles, the overall fit becomes worse.  In
itself, this mismatch is not surprising, when we remember that the
overall shape is {\em not} well described by a square root behavior
in the first place (see Table~\ref{tab:fit_vs_cut}).

\begin{figure}
\begin{center}
\epsfxsize=0.6\textwidth
\epsffile{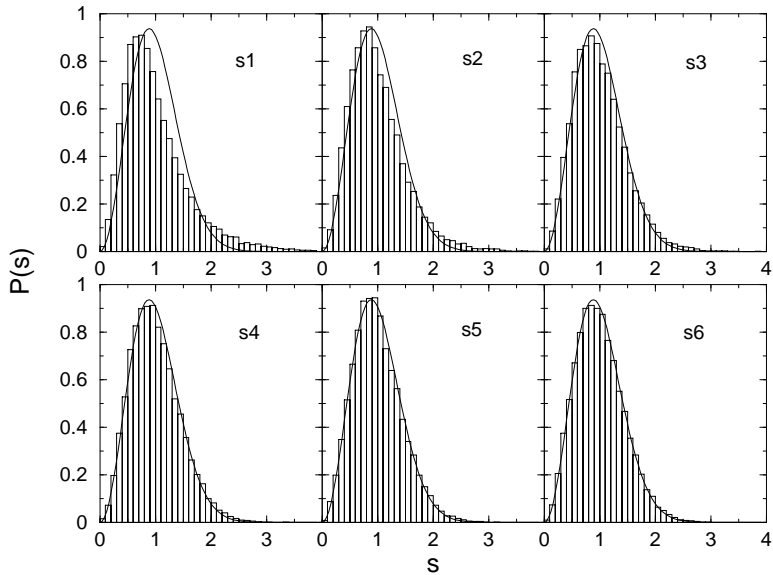}
\end{center}
\vspace*{-2mm}
\caption[a]{Distributions of the unfolded level spacings between the
first few eigenvalues in the dynamical case, compared to the Wigner surmise.
As in the quenched case, we find a small disagreement for the first
eigenvalues. But in contrast to the quenched case the agreement sets
in after almost just the 2nd eigenvalue, at the same volume.}
\label{fig:d_level_space}
\end{figure}

\noi 
In the quenched simulations we saw that correlations between the
first 5 or 6 eigenvalues were not very accurately given by Random Matrix
Theory.  In
Fig.~\ref{fig:d_level_space} the distribution of $s_i$, defined in
eq.~(\ref{eq:level_space}), is shown for a dynamical simulation with
$\beta=5.2$ , $am_q=0.025$ on a $8^3\times4$ lattice.  
Here we see, on the same lattice volume, 
a clear deviation from the Wigner surmise
only for the first two or three spacings, $s_1$, $s_2$ and $s_3$. Again,
comparisons can only be made at equal volumes, as larger volumes imply
more eigenvalues in the region around the soft edge where correlations
are poorly described by Random Matrix Theory.

\noi
There is an interesting physical consequence of a genuine gap in
the Dirac eigenvalue spectrum. Consider the difference between the
($\pi$) susceptibility $\chi_P$ and the susceptibility of its scalar 
partner ($a_0$), which we denote by $\chi_S$ \cite{B,C}. In a manner similar
to the Banks-Casher formula for the chiral condensate one finds that
this difference can be written
\beq
\omega ~=~ 4m^2 \int_0^{\infty} \! d\lambda ~\frac{\rho(\lambda;m)}{
(\lambda^2+ m^2)^2} ~,
\eeq
where the spectral density $\rho(\lambda;m)$ includes the zero modes
due to topology as well. In the infinite-volume limit this contribution
from exact zero modes should vanish, and we should be left with the
integral over non-zero modes. If the spectral density has a gap around
the origin this means that $\omega$ vanishes in the chiral limit. Because
axial U(1) rotates $\chi_P$ into $\chi_S$ and vice versa, this would
imply a restoration of axial U(1) at these high temperatures \cite{B,C}.
Conversely, if one believes that this axial U(1) symmetry is {\em not}
restored at high temperature, one gets constraints on the behavior of
the spectral density of the Dirac operator near the origin. For a power-law
behavior near $\lambda \sim 0$ a non-vanishing $\omega$ in the
infinite-volume chiral limit can only be supported for $\rho(\lambda,0)
\sim \lambda^{\alpha}$ with $\alpha \leq 1$, and in fact $\omega$ would
only remain finite in that limit if $\alpha = 1$.
 
\noi
Of course, the fact that we appear to see a gap in the eigenvalue
spectrum with staggered fermions at this particular coupling does not
imply that this gap persists in physical units as we take the continuum
limit. If not, we are here studying a pure lattice artifact. The only
way to test this is to study the shift in the apparent cut-off
eigenvalue $\lambda_0$ as we change the lattice spacing (this is, however,
far beyond what we can do at present). A very different uncertainty
comes from the fact that we are working with fermions that are almost
insensitive to topology at the couplings and volumes available to us.
This means that there are some would-be zero modes mixed up with our
regular eigenvalues near the origin. At zero temperature we found that
these would-be zero modes somewhat surprisingly behaved as the non-zero
eigenvalues at realistic couplings and volumes \cite{DHNR}. But there
is no guarantee that this is the case at finite temperature. This means
that the small tail extending to zero in our quenched simulations, and
some of the smallest eigenvalues in the simulations with dynamical
fermions may be due to these would-be zero modes. In particular, the
disagreement with Airy-function behavior very close to the soft edge
may be partly due to these would-be zero modes.

\section{Random Matrix Theory}
\label{RMT}

We have just shown our lattice gauge theory data for the spectrum of the
Dirac operator below, around, and above $T_c$, and compared it with
some of the analytical formulas of Random Matrix Theory in the limit
where the size of these matrices $N$ goes to infinity. It is of interest
to see how a simple {\em model} of a chiral phase transition in Random
Matrix Theory behaves at finite $N$, as some external parameter (mimicking
temperature $T$) is changed. The model we shall focus on here
was proposed by Jackson and Verbaarschot in the first of Ref.~\cite{JV}
and has also been studied in Ref.~\cite{Tilo}. 
It is based on a modified chiral ensemble
of $N\times N$ complex matrices $W$, with partition function 
\beq
{\cal Z} ~=~ \int\! dW \prod_{f=1}^{N_{f}}\det(M - im_f) \exp\left[-N
Tr(WW^{\dagger})\right] \label{RMM} 
\eeq
in a sector of topological charge zero. Here  
\beq
M ~=~ \left( \begin{array}{cc}
              0 & W^{\dagger} + T\\
              W + T & 0
              \end{array}
      \right)
\eeq
is a $2N\times 2N$ 
block hermitian matrix, and the external deterministic parameter $T$
is playing a r\^{o}le reminiscent of temperature. In the 
normalization chosen here, a continuous ``phase transition'' occurs at
$T_c = 1$, where the spectral density $\rho(\lambda)$ of the eigenvalues
of the random matrices just reaches zero \cite{JV}. For $T > T_c$ the
spectrum becomes two-banded, with a gap surrounding zero. The global shape 
of $\rho(\lambda)$ in this model will be very
different from the actual (macroscopic) Dirac operator spectral density. But
the interesting feature lies in having here a simple model which qualitatively
seems to describe some of the observed behavior of the staggered Dirac operator
with $N_f=4$, in particular the apparent
gap in the density for $T > T_c$.\footnote{Many other details do not match
at all. For instance the ``phase transition'' in the model (\ref{RMM}) is
continuous, in contrast to the finite-temperature phase transition in the 
massless $N_f=4$ theory, which is believed to be of first order.} 

\noi
It is very simple to do quenched ($i.e$, $N_f=0$) 
numerical simulations of the above 
Random Matrix model, as it just corresponds to generating an ensemble
of complex matrices with Gaussian weight. Such matrices are ``maximally
random'' in that their matrix elements are independently of Gaussian 
distribution. This
allows us to choose very large random matrices numerically, and then 
finding the eigenvalues of the hermitian matrix $M$. In this way we have
studied the detailed behavior of the finite-$N$ Random Matrix model
(\ref{RMM}) just below $T_c$, at $T_c$, and just above $T_c$ (where the
two bands separate as the gap develops).\footnote{Similar numerical
studies have been performed by K. Splittorff, M.Sc. thesis, The Niels Bohr 
Institute 1999 (unpublished). Simulations of the
macroscopic spectral density in this model can also be found in the original
paper by Jackson and Verbaarschot \cite{JV}.}   

\begin{figure}
\begin{center}
\epsfxsize=0.6\textwidth
\epsffile{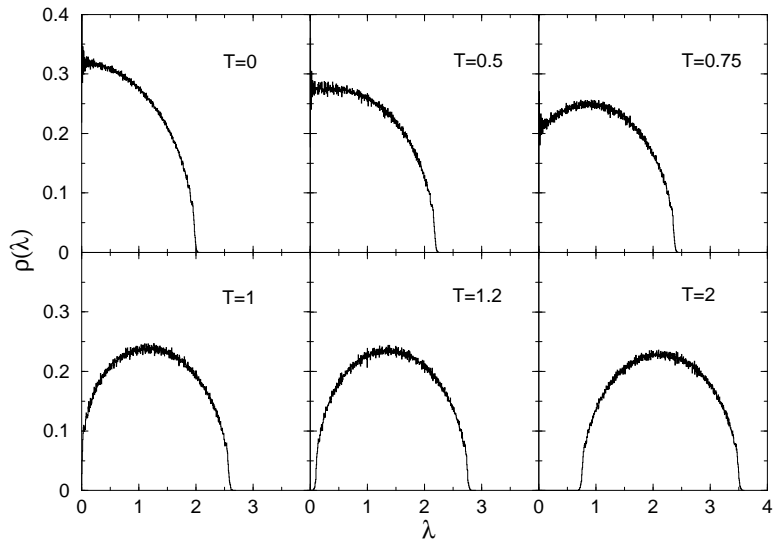}
\end{center}
\vspace*{-2mm}
\caption{Full spectra of $200\times200$ matrices at $T=0, 0.5, 0.75, 1.0,
1.2, 2.0$. In this simple model a gap indeed develops around 
$\lambda = 0$ at $T=1$.}
\label{fig:diff_macroscopic}
\end{figure}

\begin{figure}
\begin{center}
\epsfxsize=0.6\textwidth
\epsffile{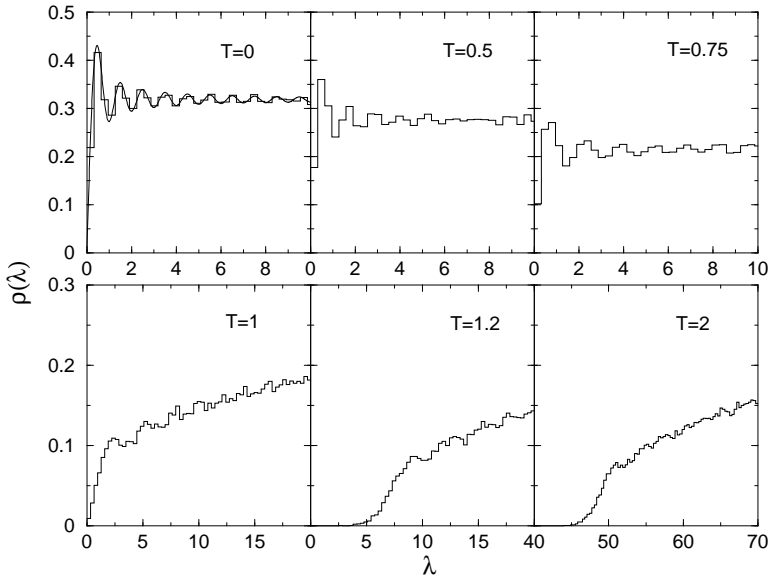}
\end{center}
\vspace*{-2mm}
\caption{A blowup of the graphs in Fig.~\ref{fig:diff_macroscopic}. The 
microscopic spectrum changes smoothly from Bessel-type behavior to
Airy-type behavior, going through an intermediate ``critical'' 
distribution at $T=1$ corresponding to a macroscopic power-law behavior of
type $\rho(\lambda) \sim \lambda^{1/3}$. The exact analytical curve for the
microscopic spectral density at $T=1$ has been computed in
Ref.~\protect\cite{Z}.}
\label{fig:diff_microscopic}
\end{figure}

\noi
We show some of our numerical results in Figs.~\ref{fig:diff_macroscopic} 
and \ref{fig:diff_microscopic}. First, in 
Fig.~\ref{fig:diff_macroscopic} we display a sequence of the macroscopic 
spectral density with increasing magnitude of the
parameter $T$: $T= 0, 0.5, 0.75, 1.0, 1.2, 2.0$. The plots were made by
diagonalizing 10000 $200\times200$ matrices. At $T=0$ this macroscopic
Random Matrix Theory spectrum is just the Wigner semi-circle law (we display
only the $\lambda > 0$ part):
\beq
\rho(\lambda,T=0) = \frac{1}{2\pi}\sqrt{4 - \lambda^2} ~.
\eeq

\begin{figure}
\begin{center}
\epsfxsize=0.6\textwidth
\epsffile{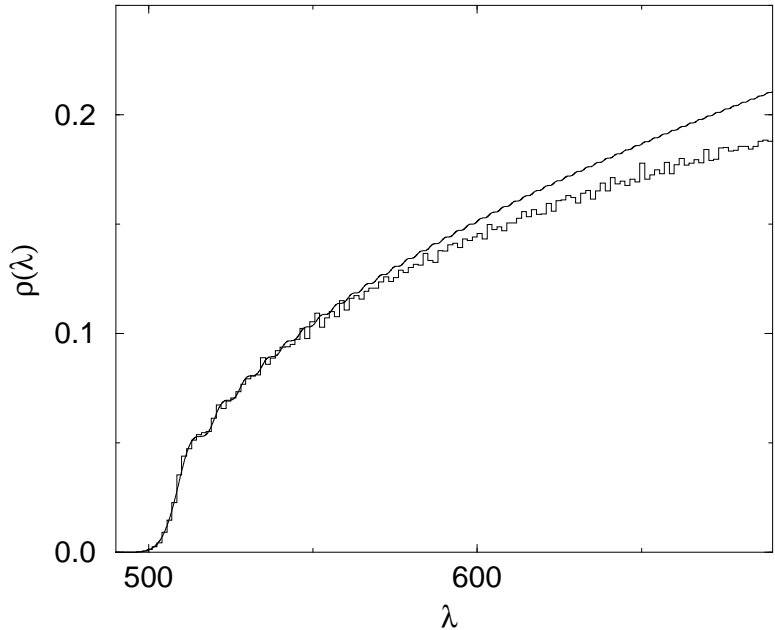}
\end{center}
\vspace*{-2mm}
\caption{A blowup of a $T=3$ spectrum compared with the Airy-prediction.
There is nice agreement near the edge.}
\label{fig:matrix_airy_compare}
\end{figure}

\noi
As $T$ increases, a dip in the spectral density around $\lambda = 0$ slowly
develops, and it subsequently turns into a gap. Of course, this macroscopic
Random Matrix Theory spectrum is totally unlike the macroscopic
Dirac operator spectrum. But it is interesting to zoom in on the microscopic
behavior of the Random Matrix spectral density at the soft edge of the
gap. In this way, by blowing up the scale of the smallest eigenvalues,
we obtain the plots in Fig.~\ref{fig:diff_microscopic} for the same 
parameter values of $T$ as above.
One sees clearly how the universal Bessel-kernel behavior below $T_c$
turns into the also universal Airy-kernel above $T_c$. We note that it is
has been shown in Ref.~\cite{JV} that the massless microscopic spectral
density of the above Random Matrix model has precisely the usual 
zero-temperature form,
\beq
\rho_s(\zeta(T)) ~=~ \frac{\zeta(T)}{2}\left[J_{N_{f}}(\zeta(T))^2
- J_{N_{f}-1}(\zeta(T))J_{N_{f}+1}(\zeta(T))\right] ~, 
\label{rhobessel}
\eeq
where $\zeta(T)$ is simply the eigenvalues rescaled by the (T-dependent) 
infinite-volume spectral density at the origin $\rho(0,T)$:
\beq
\zeta(T) ~=~ \lambda 2\pi N\rho(0,T) ~.
\eeq
In this model $\rho(0)$ approaches zero with a mean-field type of
behavior \cite{JV}:
\beq
\rho(0,T) ~=~ \rho(0,0)\sqrt{1 - T^2} ~.
\eeq
For $T$ bigger than $T_c$, but still close to it, we find, as expected,
a deformation of the Airy-kernel. In fact, the microscopic behavior
there smoothly interpolates between the Bessel-form and the Airy-form.
The peaks corresponding to individual eigenvalues from the Bessel-function
behavior below $T_c$ gradually smoothen out to become the inflection points
in the spectral density of the Airy-kind as the soft edge moves away from
the origin.
To illustrate how accurately one reproduces the Airy-behavior
in this kind of simulations, we show in 
Fig.~\ref{fig:matrix_airy_compare} the soft edge prediction appropriately 
scaled to fit simulation data at T=3.
The Airy-behavior is perfectly reproduced close to the edge, but with 
deviations after the first few eigenvalues(wiggles).  
The deviation is presumably caused by the limited number of eigenvalues
at $N=300$, and when $N$ is increased we expect the fit to improve.

\section {Conclusions}

The purpose of this study has been to find the extent to which Random
Matrix Theory may be able to describe low-lying eigenvalue distributions and
correlations between low-lying
eigenvalues of the staggered Dirac operator at finite
temperature. We have also been interested in testing the quenched theory
with staggered fermions in the light of recent results with overlap
fermions \cite{EHKN_fT}, which indicated that the chiral 
finite-temperature phase transition in the quenched theory may be more
subtle than previously expected. In the quenched case we do not see the
accumulation of small Dirac operator eigenvalues around $\lambda=0$ that
was observed with overlap fermions. This is not entirely surprising in
view of the insensitivity of staggered fermions to gauge field topology
at these lattice couplings and lattice volumes \cite{DHNR,FHL}. With 
staggered fermions we do observe a strong depletion of eigenvalues
near the origin once the temperature $T$ reaches the pure gauge theory
deconfinement phase transition temperature $T_c$. This is in agreement with 
the conventional picture that chiral symmetry is restored in the quenched 
theory with staggered fermions at precisely the deconfinement phase
transition.

\noi
On the other hand, we find a clear difference in the behavior of the smallest
Dirac operator eigenvalues in the quenched theory and the theory with
genuine, dynamical, fermions. In the quenched case a small, volume
independent tail of the eigenvalue distribution extends to $\lambda=0$
while in the full theory the tail does not reach the origin at the couplings
we have investigated. While the
bulk of the eigenvalue distribution near the (soft) edge is roughly
compatible with a square root behavior, the tail of small eigenvalues
is considerably larger than the Airy function behavior that Random
Matrix Theory would predict.
Physically, a genuine gap in physical units in the eigenvalue spectrum 
above $T_c$ would imply the restoration of axial U(1) symmetry at these
high temperatures. A very likely scenario is therefore that the
apparent gap found with staggered fermions at finite bare coupling $\beta$
shrinks to zero in physical units as the continuum limit is reached.
However, an investigation of whether this is indeed the case
is much beyond the scope of the present paper.  

\noi
{\sc Acknowledgments:}~ We thank K. Splittorff and J. Verbaarschot
for discussions.
The work of P.H.D. and K.R. has been partially 
supported by EU TMR grant no. ERBFMRXCT97-0122, and the work of U.M.H.
has been supported in part by DOE contracts DE-FG05-85ER250000 and
DE-FG05-96ER40979. In addition, P.H.D. and U.M.H. acknowledge the financial
support of NATO Science Collaborative Research Grant no. CRG 971487.

\end{document}